25th International Conference on Production Research Manufacturing Innovation:
Cyber Physical Manufacturing
August 9-14, 2019 | Chicago, Illinois (USA)

# Application of Clustering Analysis for Investigation of Food Accessibility

Rahul Srinivas Sucharitha[a]*, Seokcheon Lee[a]

*aPurdue University, School of Industrial Engineering, West Lafayette 47906, USA*

**Abstract**

Access to food assistance programs such as food pantries and food banks needs focus in order to mitigate food insecurity. Accessibility to the food assistance programs is impacted by demographics of the population and geography of the location. It hence becomes imperative to define and identify food assistance deserts (Under-served areas) within a given region to find out the ways to improve the accessibility of food. Food banks, the supplier of food to the food agencies serving the people, can manage its resources more efficiently by targeting the food assistance deserts and increase the food supply in those regions. This paper will examine the characteristics and structure of the food assistance network in the region of Ohio by presenting the possible reasons of food insecurity in this region and identify areas wherein food agencies are needed or may not be needed. Gaussian Mixture Model (GMM) clustering technique is employed to identify the possible reasons and address this problem of food accessibility.

*Keywords:* Food Insecurity; Data Analytics; Data Mining

## 1. Introduction

Food insecurity is defined as the ambiguity or lack of capability to obtain nutritionally adequate and safe food in ways that are socially acceptable (e.g., without resorting to stealing, scavenging, or other kinds of coping strategies)

* Corresponding author. Tel.: 1-765-418-5913.
  *E-mail address:* sriniv19@purdue.edu

[1]. This condition is pervasive, affecting populations all over the world. In the United States, it impacts every community with food insecurity existing in every county in America [1]. The United States Department of Agriculture (USDA) has estimated that as of 2017, close to 40 million people have been living in food insecure households with 6.5 million of them constituting children [2].

In the United States, there are a variety of involvements linking collective effort between government, public, and private entities for people suffering from inadequate access to food. One of the largest national non-profit hunger relief organization tackling hunger and food insecurity in the country is Feeding America [3].

Feeding America offers food and aid to the food insecure by having a nation-wide network of around 200 food banks and around 60,000 food pantries and meal programs (food agencies in the supply chain). Food banks attain donated food from food donors such as national food manufacturers and retailers. The donated food is transported back to the food bank using either rented or owned trucks. These donated foods are processed in the food bank for quality purposes. The trucks are then used to distribute the quality-inspected foods to the food agencies based on their accessibility. The people receive the donated foods from the food agencies.

In general, food banks operate as wholesalers of excess food. They solicit and stock bulk donations from community, government and private donors, stockpile and warehouse goods and finally distribute to the food agencies [4]. Considering the scenario is donation-driven, the supply chain is complicated as there is a dilemma in matching the supply with the demand. For example, without knowing in advance the frequency, quality and number of donated items, the supply is highly uncertain. Similarly, without estimating the need of food considering the factors leading to it (poverty, distance, etc.), the demand is highly uncertain as well. Hence, the goal of a donation-driven supply chain such as the food bank supply chain is to maximize relief for the people in need while minimizing food waste. We fill this gap by explicitly studying the nature of the families visiting the food agencies based on demographical information, and the distance between each family and their assigned food agency to locate the food assistance deserts in the region of Ohio by demonstrating a clustering technique and observe the geographic and demographic intricacies of the given region in detail.

The intent of this paper is multi-fold. The foremost aim is to notice the possible factors affecting food insecurity in a specific food bank supply chain with a particular emphasis on the demand side of the supply chain structure. A number of studies has been done with respect to understanding the supply-side of the food bank supply chain [5], [6]. However, to the best of our knowledge, the application of unsupervised machine learning techniques, specifically probabilistic models such as the mixture models to observe the demand uncertainty has not been addressed. Our study has a particular merit as it is important for non-profit organizations to leverage knowledge and technology to estimate the demand based on several complex indicators. With a good understanding of the relationship between the families getting serviced by the food agencies and certain demographic factors, food banks can arrange their personnel with improving the food accessibility and equitable distribution of food in their assigned regions, thereby improving the efficiency in their operations.

In this paper, we use the data provided by the Greater Cleveland Food Bank (GCFB) and analyse it and combine it with the demographic data provided by the USDA. The USDA data set provides the household income of each of the counties in Ohio. The subset of the data provided by GCFB consists of latitudinal and longitudinal locations of the families that are getting serviced by the food agencies (their location provided as well) and the characteristics of the family members in each family visiting the food agency. This dataset is also consisting of the census-related data as well. Using a probabilistic model such as Gaussian Mixture Model (GMM) as a clustering approach, the whole dataset can be analysed clearly despite the varying sizes and densities of each variable, thereby identifying regions within each cluster that lack food agencies near families that are in dire need and vice versa.

Similar to [7], the main research question that this paper attempts to answer is to observe if the food agencies are serving the people it intends to serve sufficiently or sparsely. Along with this question, this paper also attempts to observe the demographics of the people visiting the food agencies and if the food agencies are providing the optimum coverage of donated foods based on the census-related data.

The remainder of the paper is organized as follows. Section 2 provides a brief literature review. Section 3 presents the data. The proposed methodology is presented in section 4. Results are provided in section 5. Finally, section 6 concludes the paper and presents the potential future work.

## 2. Literature Review

Poverty and food insecurity are considered as long-term humanitarian issues, requiring the need to consider the necessity for equitable distribution of resources [8]. There has been extensive research done in the area of humanitarian logistics with importance towards the challenges experienced by non-profit food assistance programs like food banks and food pantries as explained by [9]. Food bank supply chains align with the description of humanitarian supply chains by responding to the disaster of food insecurity which can occur unexpectedly (i.e., job loss, natural calamity, etc.) or slowly (i.e., poverty) [10].The research presented in this paper aids in finding the different possible factors affecting food insecurity, thereby facilitating in improving the accessibility of food and equitable distribution of resources to the people in need. According to [11], there are innumerable ways of studying the food accessibility for people. One among the ways is studying food deserts. Food deserts are regions lacking in sources of healthy foods. This concept of food accessibility will be implemented in our paper.

There has been a fair amount of work that studies the subject of food bank supply chain. Mathematical models were introduced by [5], [6] to enable the equitable and effective distribution of food donations to the people in need. Linear programming models were formulated with the maximization of effectiveness and an equity constraint developed to solve the distribution of donated foods. Deterministic network-flow models were used to reduce the quantity of undistributed food. Several logistical issues that are being faced by the food banks have also been taken into consideration by considering the transportation schedules and permitting food banks to gather food from the limited food donors and finally transporting it to the food agencies [9]. In this paper, Food Delivery Points (FDPs) were proposed. FDPs were obtained by locating them using geographical information. The vehicle capacity and food degeneration constraints were considered during the assignment of food agencies to the respective FDPs. Using the optimal assignment, schedules were created that reflects the collection and distribution of donated food. However, these mathematical models do not investigate the varying demands of the various food agencies from where the accessibility of food is studied.

Demand of food from the food pantries and other food agencies has been taken as a deterministic value in previous non-profit based supply chain literature. On the hindsight, the demand of food is dynamic and uncertain in nature. Obtaining a way to observe the demand of food from the food agencies that the food banks are assigned to aids in understanding the possible issues arising out of food insecurity. According to [10], demand comes in the form of supplies and people for non-profit organizations and in the form of products and services for for-profit organizations, with varying demand patterns for both. [12] developed a food distribution policy using suitable welfare and poverty indices and functions to ensure an equitable and fair distribution of donated foods as per the varying demands and requirements of the people. However, the factors causing the demand or food insecurity had not been considered thereby taking several assumptions in their simulation study. Also, the model developed was suitable only for a single day period. Supply of the donated food can be done based on suitable forecasting procedures. Supply based on this kind of non-profit supply chain would be mainly dealing with guaranteeing enough inventory for the demand and reviewing the changing nature of the supply of the different types of donated foods.

In terms of implementing suitable data mining techniques, there has been relevant literature discussing the role of these techniques in the estimation of future demand using historical data in various domains. In terms of using forecasting techniques in estimating the dynamics of food donation and distribution process, [13] performed comprehensive numerical studies to quantify the extent of uncertainty in terms of the food donors, the food products, and the supply chain structure. Several predictive models were developed to estimate the quantity of in-kind donations. Predictive modelling techniques like multiple linear regression, structural equation modelling and neural networks were used in [14] to study the dynamics of food donation behavior thereby, predicting the daily average food donated by different food providers in the given region. However, the lack of statistical analysis techniques used for the study of the demand dynamics in the food bank supply chain has been evident and has been mentioned as an important challenge from the non-profit supply chain perspective [1]. Recent work addressing this issue has aided in the better understanding for the mitigation of food insecurity. [7] implemented K-means clustering to identify the food assistance deserts, a term coined by [11] while analyzing the spatial inequality existing between the rural and urban areas in access to food agencies. The results obtained from the analysis in [7] was useful in targeting the underserved areas in the given region. However, considering the dataset used consists of variables of different sizes and density, the affected families and certain traits could have been hidden and unobserved keeping in mind the lack of flexibility in a clustering technique such as K-means clustering [15].

In this paper, we address the issue of food insecurity in Ohio by analyzing the food agency service data provided by GCFB and combining the demographic data provided by the USDA and implementing GMM clustering method to the combined data based on the distances between the family visiting the food agency and the food agency serving them and observe the factors leading to food insecurity and provide ways to increase the accessibility of food.

## 3. Experimental Data

The data obtained from GCFB provides the service data of all the food agencies that the food bank distributes the donated food towards. GCFB distributes to food agencies situated in 47 counties in Ohio. The raw data is a big dataset consisting of a year's worth of family level service data with 600,000+ data points where each row represents one service to one family. It includes the latitudinal and longitudinal location points of each family and the represented food agency. Hence, we had to calculate the distance between each family and its assigned food agency to observe if they are located at an acceptable distance or not. The dataset consisted of the census tract and census block details of each family thereby aligning the distance measured as well and since the region under study is predominantly an urban area, the threshold level of distance to be considered as a food assistance desert or low food accessibility is taken to be a 1-mile demarcation [16]. The GCFB raw dataset also provided details of the number of children, seniors and adults in every family that is being served by their food agencies.

USDA also provides information of the household income and median household income of the counties at the census tract level. This data was obtained to study the low- and high-income population for the region of observation.

## 4. Proposed Methodology

The methodology followed for the clustering analysis is depicted in figure 1.

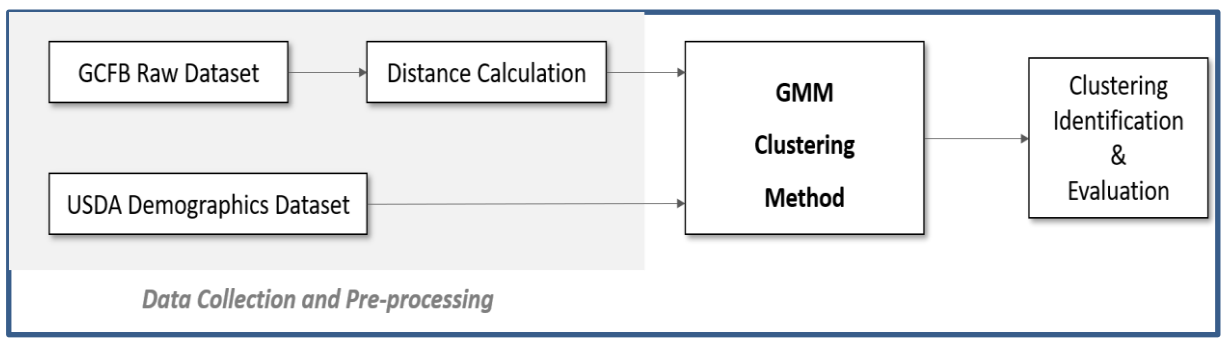

Fig. 1. Proposed Methodology

In the data collection and pre-processing stage, both the GCFB raw dataset and the USDA income-related dataset will be combined. The distance between each census tract and the assigned food agency and the distance between each family to the assigned food agency will be calculated and saved a variable. After this step, clustering using GMM method is done based on their distances and the demographics of each cluster is studied and observed.

## 5. Results and Discussion

Gaussian Mixture Model clustering is performed on the cleaned dataset using an open source software called R. The dataset is fed into the selected clustering model. GMM employs the Expectation- Maximization (EM) algorithm that only guarantees that we land to a local optimal point, but it does not guarantee that this local optima is also the global one [15].

Since we need to evaluate the clustering process and observe that the number of clusters generated by GMM is in fact the appropriate number, we explore two techniques- The Silhouette score [7] and the Bayesian Information Criterion (BIC) [15]. Using these two techniques, we can observe the right number of clusters which can be discerned from the dataset. Additionally, BIC is computed over several models and a range of values for number of clusters.

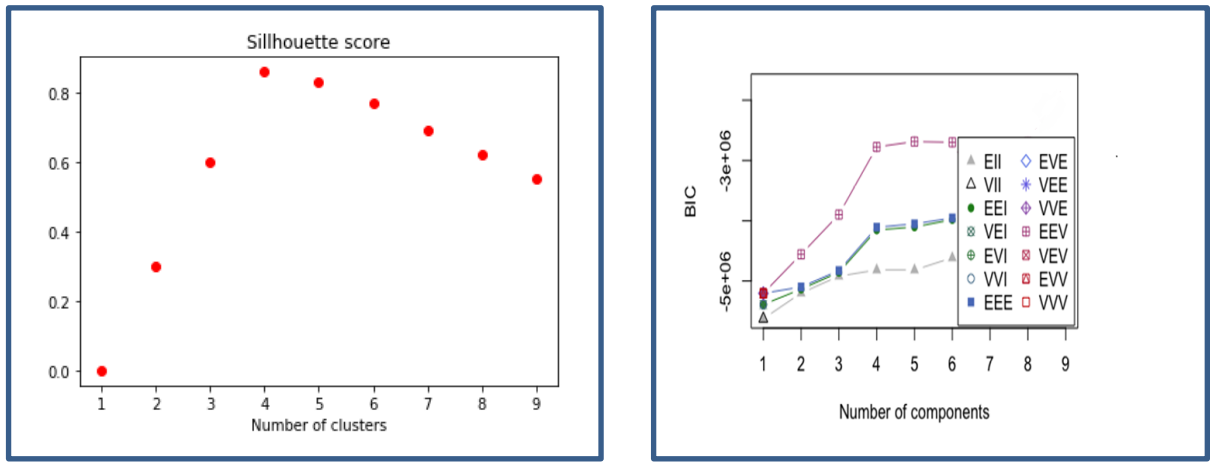

Fig. 2. (a) Average silhouette scores for different number of clusters; (b) Plot of BIC values for a variety of models and a range of number of clusters

From Figure 2 we observe that the average silhouette score is the maximum when the number of clusters is 4 and in figure 3 we observe that after a steep incline from clusters 3 to 4 it has been fairly steady when the number of clusters tallied 4. Hence, from BIC, model EEV (Ellipsoidal, equal volume, and equal shape) with 4 clusters is taken as best solution.

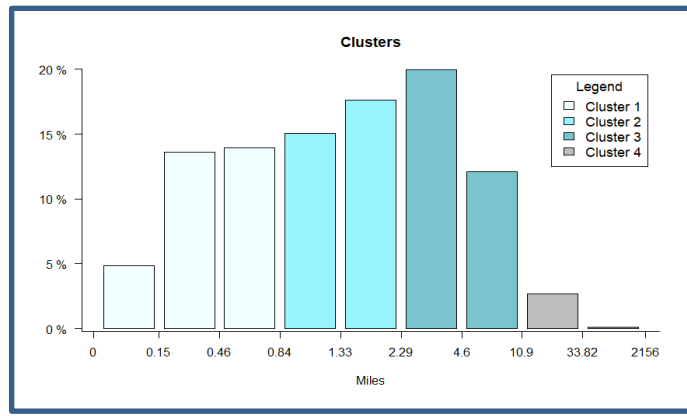

Fig. 3. Spread of distances among the clusters

The four clusters have been named based on the proximity of the families with their respective food agencies. Figure 3 provides the quantiles of the distances of every cluster. It is seen that the average distance between cluster 1 and 2 is 1.03 miles, between 2 and 3 is 3.11 miles and between 3 and 4 is 14.65 miles. These quantiles are reasonable considering that the dataset is intended for finding factors concerning food insecurity for an urban area.

Table 1. Observed values of variables per cluster

| Variables | Cluster 1 - Very Nearby | Cluster 2 - Nearby | Cluster 3 - Far Away | Cluster 4 - Very Far Away | Total |
|---|---|---|---|---|---|
| Number of families | 197,844 | 199,475 | 195,081 | 17,009 | 609,409 |
| Number of adults | 221,523 | 244,783 | 245,107 | 23,431 | 734,844 |
| Number of children | 122,884 | 154,202 | 152,667 | 13,967 | 443,720 |
| Number of seniors | 120,513 | 127,827 | 130,792 | 10,417 | 389,549 |
| Number of people | 464,920 | 526,812 | 528,566 | 47,815 | 1,568,113 |
| Average number of adults in family | 1.12 | 1.23 | 1.26 | 1.38 | 1.21 |
| Average number of children in family | 0.62 | 0.77 | 0.78 | 0.82 | 0.73 |
| Average number of seniors in family | 0.61 | 0.64 | 0.67 | 0.61 | 0.64 |
| Average number of people in family | 2.35 | 2.64 | 2.71 | 2.81 | 2.57 |
| Number of tracts | 464 | 545 | 654 | 884 | 943 |
| Average Distance (miles) | 0.42 | 1.45 | 4.63 | 19.47 | 2.64 |
| Coverage | 32.5% | 32.7% | 32.0% | 2.8% | 100.0% |
| Pct of Poor People | 92.8% | 91.2% | 84.3% | 74.7% | 89.1% |
| Pct of Rich People | 7.2% | 8.8% | 15.7% | 25.2% | 10.9% |

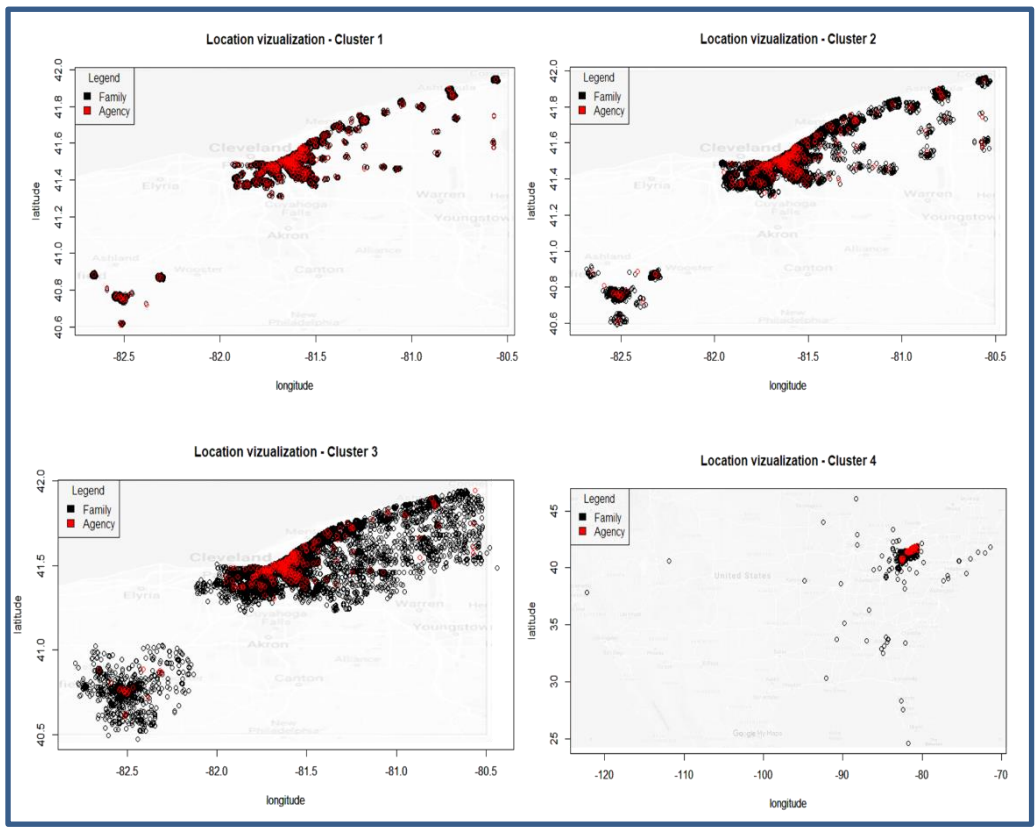

Fig. 4. Distance between families and agencies in each cluster

In Table 1, we observe that the further away the families are from the food agencies, the more people there are in the family. This holds true for children and especially adults, where as not so much for the seniors. From this table, we can interpret the number of children and seniors having low access toward food resources. It is seen that around 13,967 children live very far away from the food agencies while 122884 children live less than 0.42 miles away from the nearest food assistance. Also, the number of tracts increases as the distances from the food assistance increases. This makes sense since they are more scattered over the urban areas. Regarding the income of families in each cluster, we can see that most families (89.1%) live in tracts that are considered poor (tracts are considered poor if their average household income is less than Ohio's median income, according to [16]). It is also seen that around 11% of the families are living in tracts having high income levels.

The clustering also led to the observation that the further away the families live from the access of food, the more probable it is that they live in a tract having population receiving high income. While this observation makes sense considering that richer families do not require food assistance, it is important to observe the poor families in these regions that do not have access to food. The specific tracts and locations of families are not presented considering the huge amount of data provided for this region, but they are very well known and can be easily identified in the given dataset. Moreover, similar observations cannot be said for other regions around the country and hence, the implementation of these tools to other regions will aid in the improving the overall accessibility to food and equity of distribution of resources. The coverage is the percentage of families that are covered by at least one food agency within a mile. The only study that we know of that performed a similar kind of analysis is the study by [7]. But they implemented census tracts which is more general than the actual locations of the families. It can be seen that to increase the coverage of supply of food assistance to the poor people located very far away, some of the food agencies should be moved from cluster 1 to cluster 4 where there are areas with less coverage and people with low income.

Fig. 4. shows the spread of the agencies and families in each cluster. The latitude and longitude information were used to plot the corresponding graph for each cluster on the map of the Ohio region. It is clearly visible that the distance increases in each cluster with the average distances of each cluster mentioned in Table 1.

## 6. Conclusion

Our research analyzed the demand uncertainty in the context of food donations for a non-profit hunger relief organization. Our work contributes to the nascent humanitarian supply chain literature by providing clustering techniques as a means to comprehensively characterize the regions in low access of food assistance and finding ways to increase their access to food. With this information, GCFB can manage and distribute their food resources to the intended population in an efficient and equitable manner by targeting the regions of food assistance deserts, increasing the coverage in regions of people receiving low income and located far away from the source of food assistance.

A future direction in this research is the development of predictive models to identify the intended demand in the given region using the current dataset to optimize the food bank allocation of food in an efficient and effective manner. The developed predictive model results can be used as input parameters to mathematical models developed to improve the equitable distribution of donated foods to the people in need.

### Acknowledgements

We would like to express our sincere appreciation to Phil Trimble from the Greater Cleveland Food Bank for his valuable input into this project.